\documentclass{elsart}

\usepackage{natbib}

\usepackage{epsfig}

\usepackage{amssymb}

\begin{document}

\begin{frontmatter}
\title{Neutrino cooling rates due to $^{54,55,56}$Fe for presupernova evolution of massive stars}

\author{Jameel-Un Nabi\thanksref{footnote2}}
\address{Faculty of Engineering Sciences, GIK Institute of Engineering
Sciences and Technology, Topi 23640, Khyber Pakhtunkhwa, Pakistan}
\thanks[footnote2]{Current
Address: The Abdus Salam ICTP, Strada Costiera 11, 34014, Trieste, Italy}
\ead{jameel@giki.edu.pk, jnabi@ictp.it\\ Telephone: 0092-938-271858\\Fax:
0092-938-271890}

\begin{abstract}
Accurate estimate of neutrino energy loss rates are needed for the
study of the late stages of the stellar evolution, in particular for
cooling of neutron stars and white dwarfs. The energy spectra of
neutrinos and antineutrinos arriving at the Earth can also provide
useful information on the primary neutrino fluxes as well as
neutrino mixing scenario (it is to be noted that these supernova
neutrinos are emitted from a much later stage in stellar evolution
than that considered in this manuscript). Proton-neutron
quasi-particle random phase approximation (pn-QRPA) theory has
recently being used for a microscopic calculation of stellar weak
interaction rates of iron isotopes with success. Here I present the
detailed calculation of neutrino and antineutrino cooling rates due
to key iron isotopes in stellar matter using the pn-QRPA theory. The
rates are calculated on a fine grid of temperature-density scale
suitable for core-collapse simulators. The calculated rates are
compared against earlier calculations. The neutrino cooling rates
due to isotopes of iron are in overall good agreement with the rates
calculated using the large-scale shell model. During the
presupernova evolution of massive stars, from oxygen shell burning
till around end of convective core silicon burning phases, the
calculated neutrino cooling rates due to $^{54}$Fe are three to four
times larger than the corresponding shell model rates. The Brink's
hypothesis used in previous calculations can at times lead to
erroneous results. The Brink's hypothesis assumes that the
Gamow-Teller strength distributions for all excited states are the
same. It is, however, shown by the present calculation that both the
centroid and total strength for excited states differ appreciably
from the ground state distribution. These changes in the strength
distributions of thermally populated excited states can alter the
total weak interaction rates rather significantly. The calculated
antineutrino cooling rates, due to positron capture and
$\beta$-decay of iron isotopes, are orders of magnitude smaller than
the corresponding neutrino cooling rates and can safely be neglected
specially at low temperatures and high stellar densities.
\end{abstract}

\end{frontmatter}
\parindent=0.5 cm
\section{Introduction}

Despite immense technological advancements since the time when
Colgate $\&$ White \cite{Col66} and Arnett \cite{Arn67} presented
their classical work on energy transport by neutrinos and
antineutrinos in non-rotating massive stars, the explosion
mechanism of core-collapse supernovae continues to pose challenges
for the collapse simulators throughout the globe. It is clear that
the prompt shock that follows the bounce of the core stagnates and
is not possible to cause a supernova explosion on its own. It
loses energy in disintegrating iron nuclei and through neutrino
emissions (mainly non-thermal) which are till then transparent to
the stellar matter. A few milliseconds after the bounce, the
proto-neutron star accretes mass at a few tenths of solar mass per
second. This accretion, if continued even for one second, can
change the ultimate fate of the collapsing core resulting into a
black hole. Neutrinos are the main characters in this play and
radiate around 10$\%$ of the rest mass converting the star to a
neutron star. Initially the nascent neutron star is a hot thermal
bath of dense nuclear matter, $e^{-}e^{+}$ pairs, photons and
neutrinos. Neutrinos, having the weak interaction, are most
effective in cooling and diffuse outward within a few seconds, and
eventually escape with about 99$\%$ of the released gravitational
energy. Despite the small neutrino-nucleus cross sections, the
neutrinos flux generated by the cooling of a neutron star can
produce a number of nuclear transmutations as it passes the
onion-like structured envelope surrounding the neutron star. In
the late-time neutrino heating mechanism the stalled shock can be
revived (about 1 s after the bounce) and may be driven as a
delayed explosion \cite{Bet85}.  The
2D simulations (addition of convection) performed with a Boltzmann
solver for the neutrino transport fails to convert the collapse
into an explosion \cite{Bur03}. 2D calculations carried out in the
mid -1990's  resulted in successful supernovae and revealed some
role of turbulence in the collapsing gas (e.g. \cite{Fry99}). Recently a few simulation groups (e.g.
\cite{Buras06, Bur06, Woo07} have reported successful explosions in
2D mode. However a complete understanding of the explosion mechanism is still in progress.
Additional energy sources were also sought that might transport
energy to the mantle and lead to an explosion. Few popular
sources of additional energy that were widely discussed in
literature were preheating mechanism proposed by Haxton
\cite{Hax88}, magnetic fields (e.g. see Ref. \cite{Kot04}) and rotations
(e.g. see Ref. \cite{Wal05}).

The structure of the progenitor star has a vital role to play in the
mechanism of the explosion. A lot many physical inputs are required
at the beginning of each stage of the entire simulation process
including but not limited to collapse of the core, formation,
stalling and revival of the shock wave and shock propagation. It is
highly desirable to calculate the presupernova stellar structure
with the most reliable physical data and inputs.

Neutrinos from core-collapse supernovae are unique messengers of the
microphysics of supernovae and are crucial to the life and afterlife
of supernovae. They provide information regarding the neutronization
due to electron capture, the infall phase, the formation and
propagation of the shock wave and the cooling phase. Cooling rate is
one of the crucial parameters that strongly affect the stellar
evolution. During the late stages of stellar evolution a star mainly
looses energy through neutrinos. White dwarfs and supernovae (which
are the endpoints for stars of varying masses) have both cooling
rates largely dominated by neutrino production. A cooling
proto-neutron star emits about $3 \times 10^{53}$ erg in neutrinos,
with the energy roughly equipartitioned among all species. Further
the neutrinos and antineutrinos produced as a result of nuclear
reactions are transparent to the stellar matter at presupernova
densities and therefore assist in cooling the core to a lower
entropy state. This scenario does not necessarily hold at extremely
high densities and temperatures (this would be the case for stellar
collapse where dynamical time scales become shorter than the
neutrino transport time scales) where neutrinos can become trapped
in the so-called neutrinospheres mainly due to elastic scattering
with nuclei. Prior to stellar collapse one requires an accurate
determination of neutrino energy loss rates in order to perform a
careful study of the final branches of star evolutionary tracks.
Throughout this text (anti)neutrino energy loss rates and
(anti)neutrino cooling rates are meant as the same physical phenomena and the
two terms are used interchangeably. A change in the cooling rates
particularly at the very last stages of massive star evolution could
affect the evolutionary time scale and the iron core configuration
at the onset of the explosion \cite{Esp03}. The electron capture
rates and the accompanying neutrino energy loss rates are also
required in determining the equation of state. The neutrino energy
loss rates are important input parameters in multi-dimensional
simulations of the contracting proto-neutron star. Reliable and
microscopic calculations of neutrino cooling rates and capture rates
can contribute effectively in the final outcome of these simulations
on world's fastest supercomputers.

The first-ever extensive calculation of stellar weak rates including
the capture rates, neutrino energy loss rates and decay rates for a
wide density and temperature domain was performed by Fuller, Fowler,
and Newman (FFN) \cite{Ful80}.  Later, Aufderheide et
al. \cite{Auf94} extended the FFN work for heavier nuclei with A $>$
60 and took into consideration the quenching of the GT strength
neglected by FFN. The measured data from various (p,n) and
(n,p) experiments later revealed the misplacement of the GT centroid
adopted in the parameterizations of FFN. Since then theoretical
efforts were concentrated on the microscopic calculations of
weak-interaction mediated rates of iron-regime nuclide. Large-scale
shell model (LSSM)(e.g. \cite{Lan00}) and the proton-neutron
quasiparticle random phase approximation (pn-QRPA) theory  (e.g.
\cite{Nab04}) were used extensively and with relative success for
the microscopic calculation of stellar weak rates. Monte Carlo
shell-model is an alternative to the diagonalization method and
allows calculation of nuclear properties as thermal averages (e.g.
\cite{Joh92}). However it does not allow for detailed nuclear
spectroscopy and has some restrictions in its applications for
odd-odd and odd-A nuclei.

The pn-QRPA theory is an efficient way to generate GT strength
distributions. These strength distributions constitute a primary and
nontrivial contribution to the weak-interaction mediated rates among
iron-regime nuclide. Because of the high temperatures prevailing
during the presupernova and supernova phase of a massive star, there
is a reasonable probability of occupation of parent excited states
and the total weak interaction rates have a finite contribution form
these excited states. The pn-QRPA theory allows a microscopic
state-by-state calculation of \textit{all} these partial rates and
this feature of the model greatly enhances the reliability of the
calculated rates in stellar matter. Previous calculations of stellar
weak-interaction rates (e.g. FFN and LSSM) assumed the so-called
Brink's hypothesis to approximate the contribution of partial rates
from high-lying excited states. This hypothesis assumes that the GT
strength distribution is the same as the ground-state GT strength
distribution and hence the rate contribution for each transition is
essentially the same. However it was shown in a recent pn-QRPA
calculation \cite{Nab09} that the Brinks hypothesis is a poor
approximation for key iron isotopes considered in this project.
Table 2 of Ref. \cite{Nab09} showed that in the $\beta^{-}$
direction the total strength for the first excited state of
$^{54}$Fe changed from 7.56 to 6.97, that of $^{55}$Fe increased
from 6.87 to 8.87 and finally for $^{56}$Fe decreased from 10.74 to
8.04, respectively. There were also corresponding changes in
centroids and strength in $\beta^{+}$ direction. These changes do
have an overall effect on the total rate under stellar conditions.
Further details of these calculations and the pn-QRPA model can be
found in Ref. \cite{Nab10}. The improved calculation of
weak-interaction mediated rates on iron isotopes was recently
introduced \cite{Nab09} using the pn-QRPA theory. The improvement
was attributed to a judicious choice of model parameters and
incorporation of measured deformation for the even-even isotopes of
iron. There the author was able to reproduce fairly well the
experimental centroids and the total strength distributions in both
directions (in the GT$_{+}$ direction a proton is converted to a
neutron, as in electron capture or positron decay and in the
GT$_{-}$ direction a neutron is converted to a proton, as in
positron capture or $\beta$-decay) for the even-even iron isotopes,
$^{54,56}$Fe.  This paper is devoted to a detailed analysis of the
neutrino and antineutrino energy loss rates due to $^{54,55,56}$Fe
in stellar matter. The neutrino cooling rates depend heavily on the
calculation of the associated GT strength distribution functions.
The pn-QRPA calculated GT strength functions for iron isotopes were
also introduced in Ref. \cite{Nab09} (see also \cite{Nab10} for a
discussion of the subject).  Simulation results of presupernova
evolution of massive stars do point $^{54,55,56}$Fe as key iron
isotopes whose weak-interaction mediated rates can strongly
influence the outcome of such results (e.g. see Refs. \cite{Auf94,
Heg01}) At lower temperatures and densities characteristic of the
hydrostatic phases of stellar evolution, accurate and reliable
stellar weak rates are required to determine the nucleosynthesis of
nuclear species, the overall neutrino energy loss rates which may
affect the temperature of the core (at relevant temperatures and
densities the nonthermal neutrinos are transparent to the stellar
matter), and the detailed $Y_{e}$ which becomes very important going
into the core collapse. For the later phases of silicon burning to
the collapse phase, overall neutronisation rates and neutrino
production rates become the most interesting quantities \cite{Ful80}
(during this phase the total GT$_{\pm}$ strengths become more
important rather than their distributions).

The paper is written in the following format. Section 2 describes
the essential formalism for the calculation of neutrino and
antineutrino energy loss rates using the pn-QRPA theory. I present
my calculation in Section 3 where I also compare them with earlier
calculations of neutrino energy loss rates. I summarize the main
points and conclude finally in Section 4.

\section{Formalism}
The QRPA theory considers the residual correlations among the
nucleons via one particle one hole (1p-1h) excitations in a large
model space and is an efficient way to generate GT strength
distributions.  Kar et al. \cite{Kar94} pointed out much earlier
that the quasiparticle random phase approximation (QRPA) method is
quite successful in predicting the weak interaction rates of ground
states all over the periodic table and also stressed the need to
extend these methods to non-zero temperature domains relevant to
presupernova and supernova conditions.  Nabi and
Klapdor-Kleingrothaus \cite{Nab99} later used the pn-QRPA theory in
stellar matter to calculate contributions to weak interaction rates
from parent excited states. The basic formalism of the pn-QRPA model can be found in Ref. \cite{Nab10}

The neutrino (antineutrino) energy loss rates can occur through four
different weak-interaction mediated channels: electron and positron
emissions, and, continuum electron and positron captures. The
neutrino energy loss rates were calculated using the formula

\begin{equation}
\lambda ^{^{\nu(\bar{\nu})} } _{ij} =\left[\frac{\ln 2}{D}
\right]\left[f_{ij}^{\nu} (T,\rho ,E_{f} )\right]\left[B(F)_{ij}
+\left({\raise0.7ex\hbox{$ g_{A}  $}\!\mathord{\left/ {\vphantom
{g_{A}  g_{V} }} \right.
\kern-\nulldelimiterspace}\!\lower0.7ex\hbox{$ g_{V}  $}}
\right)^{2}_{eff} B(GT)_{ij} \right].
\end{equation}

Reduced transition probabilities as well as values of constants used in Eqt.~1 can be seen from Ref. \cite{Nab10}.
A quenching factor of 0.6 was introduced in the calculation \cite{Nab09,Nab10}.
The $f_{ij}^{\nu}$ are the phase space integrals and are functions
of stellar temperature ($T$), density ($\rho$) and Fermi energy
($E_{f}$) of the electrons. They are explicitly given by
\begin{equation}
f_{ij}^{\nu} \, =\, \int _{1 }^{w_{m}}w\sqrt{w^{2} -1} (w_{m} \,
 -\, w)^{3} F(\pm Z,w)(1- G_{\mp}) dw,
\end{equation}
and by
\begin{equation}
f_{ij}^{\nu} \, =\, \int _{w_{l} }^{\infty }w\sqrt{w^{2} -1} (w_{m}
\,
 +\, w)^{3} F(\pm Z,w)G_{\mp} dw.
\end{equation}
In above equation $w$ is the total energy of the electron including
its rest mass, $w_{l}$ is the total capture threshold energy
(rest+kinetic) for positron (or electron) capture. F($ \pm$ Z,w) are
the Fermi functions and were calculated according to the procedure
adopted by Gove and Martin \cite{Gov71}. G$_{\pm}$ is the
Fermi-Dirac distribution function for positrons (electrons).
\begin{equation}
G_{+} =\left[\exp \left(\frac{E+2+E_{f} }{kT}\right)+1\right]^{-1},
\end{equation}
\begin{equation}
 G_{-} =\left[\exp \left(\frac{E-E_{f} }{kT}
 \right)+1\right]^{-1},
\end{equation}
here $E$ is the kinetic energy of the electrons and $k$ is the
Boltzmann constant.

For the decay channel Eqt. 2 was used for the calculation of phase
space integrals. Upper signs were used for the case of electron
emissions and lower signs for the case of positron emissions.
Regarding the capture channels, I used Eqt. 3 for the calculation
of phase space integrals keeping upper signs for continuum
electron captures and lower signs for continuum positron captures.
Details of the calculation of reduced transition probabilities can
also be found in Ref. \cite{Nab99a}.

The total neutrino energy loss rate per unit time per nucleus is
given by
\begin{equation}
\lambda^{\nu} =\sum _{ij}P_{i} \lambda _{ij}^{\nu},
\end{equation}
where $\lambda_{ij}^{\nu}$ is the sum of the electron capture and
positron decay rates for the transition $i \rightarrow j$ and
$P_{i}$ is the probability of occupation of parent excited states
which follows the normal Boltzmann distribution.

On the other hand the total antineutrino energy loss rate per unit
time per nucleus is given by
\begin{equation}
\lambda^{\bar{\nu}} =\sum _{ij}P_{i} \lambda _{ij}^{\bar{\nu}},
\end{equation}
where $\lambda_{ij}^{\bar{\nu}}$ is the sum of the positron
capture and electron decay rates for the transition $i \rightarrow
j$.

The summation over all initial and final states was carried out
until satisfactory convergence in the rate calculation was achieved.
The pn-QRPA theory allows a microscopic state-by-state calculation
of both sums present in Eqts. 6 and 7. This feature of the model
greatly increases the reliability of the calculated rates in stellar
matter where there exists a finite probability of occupation of
excited states.

\section{Results and comparison}

The improved calculation of GT$_{\pm}$ strength distributions for
$^{54,55,56}$Fe was introduced in Ref. \cite{Nab09} and discussed in
detail in Ref. \cite{Nab10}. There I also compared the calculated
strength functions against the measured distributions for the case
of even-even isotopes of iron. It was shown in Table 1 of Ref. \cite{Nab09}
that the comparison of the total GT strengths and centroids of
$^{54,56}$Fe with the measured data improved considerably relative
to the earlier pn-QRPA calculation \cite{Nab04}.

The summed B(GT$_{+}$) distributions for isotopes of iron were
discussed in Ref. \cite{Nab09}. Here I show the
cumulative strength distributions, B(GT$_{-}$), of $^{54,55,56}$Fe using the pn-QRPA theory.
The cumulative B(GT$_{-}$) strength distributions for the
iron isotopes are displayed in Figs. 1--3. The abscissas refer to energy in daughter cobalt isotopes. It is to be noted
that the distributions are well fragmented and extend to high
excitation energies in daughter. The corresponding GT$_{+}$
strengths are understandably lower in magnitude as compared to the
GT$_{-}$ strength. The values of the total strength functions and centroids
for the ground-state and excited state GT strength functions, B(GT$_{\pm}$), were given earlier in Ref. \cite{Nab09}.

Moving on to the calculation of neutrino energy loss rates, Figs.
4--6 depict the energy loss rates for $^{54}$Fe, $^{55}$Fe and
$^{56}$Fe, respectively. Each figure consists of four panels
depicting the calculated neutrino energy loss rates at selected
temperature and density domain. It is pertinent to mention again
that the neutrino energy loss rates contain contributions due to
electron capture \textit{and} positron decay on iron isotopes. The
upper left panel, in all figures, shows the energy loss (cooling)
rates in low-density region ($\rho [gcm^{-3}] =10^{0.5}, 10^{1.5}$
and $10^{2.5}$), the upper right in medium-low density region ($\rho
[gcm^{-3}] =10^{3.5}, 10^{4.5}$ and $10^{5.5}$), the lower left in
medium-high density region ($\rho [gcm^{-3}] =10^{6.5}, 10^{7.5}$
and $10^{8.5}$) and finally the lower right panel depicts the
calculated rates in high density region ($\rho [gcm^{-3}] =10^{9.5},
10^{10.5}$ and $10^{11}$). The neutrino energy loss rates are given
in logarithmic scales (to base 10) in units of $MeV. s^{-1}$. In the
figures T$_{9}$ gives the stellar temperature in units of $10^{9}$
K. One should note the order of magnitude differences in neutrino
energy loss rates as the stellar temperature increases  (first three
panels of Figs. 4--6). For densities $\rho [gcm^{-3}] \leq 10^{8.5}$
and low stellar temperatures (T$_{9} \leq 5$), the pn-QRPA
calculates the lowest energy loss rates due to $^{56}$Fe  and
highest due to $^{55}$Fe. It can be seen from these figures that in
the low density region the energy loss rates, as a function of
stellar temperatures, are more or less superimposed on one another.
This means that there is no appreciable change in the neutrino
energy loss rates when increasing the density by an order of
magnitude. There is a sharp exponential increase in the neutrino
energy loss rates for the low and medium-low density regions as the
stellar temperature increases up to T$_{9}$ =5. Beyond this
temperature the slope of the rates reduces drastically with
increasing density. For a given temperature the neutrino energy loss
rates increase monotonically with increasing densities. The figures
are drawn from  $5 \leq T_{9} \leq 30$ in order to span a smaller
scale and to see the differences between the rates clearly. Neutrino
cooling rates at low stellar temperatures (T$_{9} \leq 5$) are
available.

The calculated antineutrino energy loss rates contain contributions
due to positron capture \textit{and} $\beta$-decay on iron isotopes
(see Eqt. 7). Figs. 7--9 show a similar calculation for the
antineutrino energy loss rates for the selected isotopes of iron.
Again one notes that the figures are drawn from  $5 \leq T_{9} \leq
30$ due to a large variation in the magnitude of these rates as the
stellar temperature increases. The anti-neutrino rates for
temperatures below T$_{9} \leq 5$ are available and can be requested
from the author. Figs. 7--9 show that the corresponding
anti-neutrino cooling rates are orders of magnitude smaller than the
neutrino cooling rates. Specially at low temperatures and high
densities these rates can safely be neglected as compared to the
corresponding neutrino cooling rates. The rates are almost
superimposed on one another as a function of stellar densities.
However as the stellar matter moves from the medium high density
region to high density region these rates start to 'peel off' from
one another. The neutrino and antineutrino energy loss rates are
calculated on an extensive temperature-density grid point suitable
for collapse simulations and interpolation purposes. The electronic
versions of these files may be requested from the author.

An interesting query would be to know how the reported pn-QRPA
calculation compares with large-scale shell model (LSSM) calculation
\cite{Lan00} and the pioneer calculation performed by FFN
\cite{Ful80} (specially for temperature and density domains of
astrophysical interest). Of particular mention is the error in the
Lanczos-based approach employed by LSSM  and pointed by Pruet and
Fuller \cite{Pru03}. The calculated decay rates by LSSM is a
function of the number of Lanczos iterations required for
convergence and this treatment of partition functions can influence
their estimates of high-temperature $\beta$-decay rates.
Consequently LSSM rates tend to be too low at high temperatures. The
pn-QRPA calculation do not suffer from this convergence problem as
it is not Lanczos-based. Secondly, as already mention, the pn-QRPA
model does not assume Brink's hypothesis and back resonances as
employed in calculations by FFN and LSSM.

The comparisons are presented in a tabular form. Tables (1 -- 3)
show the comparison of calculated neutrino energy loss rates with
those of FFN and LSSM for $^{54}$Fe, $^{55}$Fe and $^{56}$Fe,
respectively. Here the ratios of the calculated neutrino energy loss
rates to those of LSSM, $R_{\nu}$(LSSM), and the corresponding
ratios to those of FFN, $R_{\nu}$(FFN), are presented at selected
temperature and density points. According to the study of
presupernova evolution of massive stars by Heger and collaborators
\cite{Heg01}, electron capture rates on $^{54}$Fe are important from
the oxygen shell burning phase up to the end of convective core
silicon burning phase of massive stars. For temperatures (T$_{9}
\sim$3) and densities ($\rho \sim 10^{7} gcm^{-3}$) corresponding
roughly to the oxygen shell burning phase of massive stars, the
calculated neutrino energy loss rates due to $^{54}$Fe are enhanced
by as much as a factor of three compared to LSSM results (Table 1).
The reason for this enhancement may be traced back to the three
times bigger electron capture rates on $^{54}$Fe for the
corresponding density and temperature scales using the pn-QRPA
theory \cite{Nab09} which dominate the production of these
non-thermal neutrinos (see Eqt. 6). During the later phases of
stellar evolution the pn-QRPA calculated neutrino energy loss rates
are in good agreement with the LSSM rates. FFN rates are enhanced
specially at low temperatures and densities by an order of
magnitude. FFN neglected the quenching of the total GT strength in
their rate calculation. The comparison improves at higher
temperatures and densities even though the FFN rates are still
enhanced (around a factor of three).

The electron capture rates on $^{55}$Fe are most effective during
the oxygen shell burning till around the ignition of the first stage
of convective silicon shell burning of massive stars \cite{Heg01}.
Correspondingly one should expect the most effective cooling
contribution due to $^{55}$Fe during the above mentioned phases of
stellar evolution. For the corresponding temperatures (T$_{9}
\sim$3) and densities ($\rho \sim 10^{7} gcm^{-3}$), the pn-QRPA
rates are in very good agreement with the LSSM rates (Table 2). In
fact the overall agreement is excellent at all temperatures and
densities for the case of $^{55}$Fe. The comparison is fair with FFN
rates at low temperatures and densities (though the FFN rates are
slightly bigger). At high temperatures and densities FFN rates are
much enhanced due to above mentioned reason.

The overall comparison of neutrino cooling rates with LSSM is again
good for the case of $^{56}$Fe (Table~3). It is to be noted that
both pn-QRPA theory and LSSM calculates the ground-state GT
distributions microscopically. For higher lying excited states,
pn-QRPA model again calculates the  GT strength distributions in a
microscopic fashion whereas Brink's hypothesis and back resonances
are employed in LSSM and FFN calculations. Accordingly, whenever
ground state rates command the total rate, the two calculations are
found to be in excellent agreement. For cases where excited state
partial rates influence the total rate, differences are seen between
the two calculations. FFN rates are again enhanced by a factor of
two to four as compared to pn-QRPA rates.

A similar tabular comparison of the calculation of antineutrino
energy loss rates on iron isotopes using the pn-QRPA theory against
FFN and LSSM rates is presented in Tables (4 -- 6). Here also the
ratios of the calculated antineutrino energy loss rates to those of
FFN, $R_{\bar{\nu}}$(FFN), and LSSM, $R_{\bar{\nu}}$(LSSM), are
presented at selected temperature and density points. For T$_{9} =
1$ and $\rho = 10^{11} gcm^{-3}$ all three calculations reported
cooling rates $< 10^{-100}$ MeV/s and as such determination of
ratios were not possible. The calculated anti-neutrino cooling rates
are smaller by 1--4 orders of magnitude in comparison to previous
calculations. Only at high temperatures (T$_{9} \sim$ 30) is the
pn-QRPA calculated anti-neutrino cooling rates an order of magnitude
bigger than the corresponding LSSM rates.

Table 4 presents the comparison of the calculations of antineutrino
energy loss rate for the case of $^{54}$Fe. One sees that the
comparison is fairly good against LSSM and FFN calculations at high
temperatures (T$_{9} \geq 10$). At low temperatures (T$_{9} \leq 1$)
and densities ($\rho \leq 10^{3} gcm^{-3}$) the calculated
antineutrino cooling rates are $<$ 10$^{-45}$ MeV/s. As the density
increases to $\rho = 10^{11} gcm^{-3}$ the calculated rates are
smaller than 10$^{-100}$ MeV/s. These small numbers are fragile
functions of calculated energy levels and are around four orders of
magnitude smaller than previously calculated. As the stellar
temperature increases to T$_{9} = 3$, the magnitude of the
calculated rates increase to around 10$^{-15}$ MeV/s for densities
up to $\rho = 10^{3} gcm^{-3}$, and the pn-QRPA calculated rates are
around two orders of magnitude smaller (at high stellar densities,
$\rho = 10^{11} gcm^{-3}$, the calculated cooling rates are smaller
than 10$^{-55}$ MeV/s and three orders of magnitude smaller than
previous calculations). As temperature further increases so does the
magnitude of calculated cooling rates and get in reasonable
comparison with FFN and LSSM numbers. At high temperatures, T$_{9} =
30$, the calculated rates are around a factor six bigger than LSSM
rates (it is to be recalled that LSSM calculated rates tend to be
smaller at high temperatures due to Lanczos-based approach as
pointed by Pruet and Fuller \cite{Pru03}).

The calculated anti-neutrino cooling rates due to $^{55}$Fe are 2--4
orders of magnitude smaller than previous calculations whenever they
have a very small magnitude (see Table 5). According to the study by
Heger et al. \cite{Heg01}, $^{55}$Fe is included among the top three
nuclei that increases $Y_{e}$ the most for  25 $M_{\odot}$ and 40
$M_{\odot}$ stars and are most effective during the silicon burning
phase of these massive stars. It was later shown by Nabi in Ref.
\cite{Nab10} that $\beta$-decay rates of iron isotopes were 3--5
orders of magnitude smaller than previously calculated (mainly
because of approximations like Brink's hypothesis and back
resonances employed in these calculations) and hence irrelevant for
the determination of the evolution of $Y_{e}$ during the
presupernova phases of massive stars. During this phase (T$_{9}
\sim$3, $\rho \sim 10^{7} gcm^{-3}$) the calculated corresponding
antineutrino energy loss rates are suppressed by around two orders
of magnitude compared with those of LSSM and FFN (Table 5). For
temperatures around $ 3 \leq T_{9} \leq 10$, the calculated
antineutrino energy loss rates are in very good comparison with LSSM
rates in low-density regions. An order of magnitude enhancement is
noted in calculated rates at T$_{9} \sim$ 30 compared to LSSM
numbers for reasons mentioned above. The corresponding comparison
with FFN is good at high temperatures.

The comparison of antineutrino energy rates due to $^{56}$Fe is
overall fine against LSSM rates with reasonable enhancements and
suppressions of calculated rates at different points of temperature
and density scale (Table 6). The comparison improves at higher
temperatures. The LSSM rates are enhanced up to two orders of
magnitude at low temperatures. At high temperatures the calculated
rates are enhanced around a factor of six. The rates are in good
comparison with FFN calculation at T$_{9} = 30$. Otherwise FFN rates
are enhanced by 1 --4 orders of magnitude. As mentioned earlier, the
antineutrino cooling rates are smaller than the corresponding
neutrino cooling rates by orders of magnitude and these small
numbers can change appreciably by a mere change of 0.5 MeV in phase
space calculations and are actually more reflective of the
uncertainties in calculation of the energy eigenvalues (for both
parent and daughter states)in the respective models.

\section{Summary and conclusions}

In order to understand the supernova explosion mechanism
international collaborations of astronomers and physicists are being
sought. Weak interaction mediated rates are key nuclear physics
input to simulation codes and a reliable and  microscopic
calculation of these rates (both from ground-state \textit{and}
excited states) is desirable. The pn-QRPA theory with improved model
parameters was used to calculate the (anti)neutrino energy loss
rates due to iron isotopes in stellar matter. The calculation was
performed in a luxurious model space of 7$\hbar\omega$. The
microscopic calculation of GT$_{\pm}$ strength distributions from
ground and excited states highlighted the fact that the Brink's
hypothesis and back resonances may not be good approximations to use
in calculation of stellar weak rates.

The associated energy loss rates due to weak interactions on iron
isotopes in stellar matter were calculated using the pn-QRPA theory.
Deformations of nuclei were taken into account and the calculation
took into consideration the experimental deformations for even-even
isotopes of iron ($^{54,56}$Fe). All available experimental data
were incorporated in the calculation to enhance the reliability of
cooling rates. The calculated neutrino energy loss rates due to
$^{54}$Fe are up to four times bigger than the LSSM rates during the
oxygen and silicon shell burning phases of massive stars. The
comparison with LSSM gets better for successive and supernova phases
of stellar evolution. During silicon shell burning for stars ($\sim
10 - 25M_{\odot}$) and oxygen shell burning for much heavier stars
($\sim 40M_{\odot}$) the calculated energy loss rates due to
$^{55}$Fe are in very good comparison with the LSSM rates.  The
results for neutrino energy loss rates due to $^{56}$Fe are in
overall good agreement with the corresponding LSSM numbers. For the
silicon burning phase of massive stars, the calculated antineutrino
energy loss rates are suppressed by more than two orders of
magnitude compared with the LSSM calculation and hence can be safely
neglected.

According to the study of presupernova evolution of heavy stars by
authors in Ref. \cite{Heg01} the most important period for
determining structure and lepton fraction in the core occurs during
silicon shell burning. During this decisive phase of stellar
evolution the pn-QRPA calculated energy loss rates due to $^{54}$Fe
are much enhanced as compared to the results of large-scale shell
model calculation and favor cooler cores with lower entropies. On
the other hand, during silicon burning phases of massive stars, the
antineutrino energy loss rates are suppressed by two orders of
magnitude compared to previous calculations and can be neglected.
These information may be of use for core-collapse simulators and may
contribute in the fine tuning of the temperature, entropy and the
lepton-to-baryon ratio which become very important going into
stellar collapse. The rates are calculated on an extensive
temperature-density grid point suitable for collapse simulations and
the electronic versions of these files may be requested from the
author.

\vspace{0.5 in}\textbf{Acknowledgments:} The author would like to
acknowledge the kind hospitality provided by the Abdus Salam ICTP,
Trieste, where part of this project was completed. The author also
wishes to acknowledge the support of research grant provided by the
Higher Education Commission Pakistan,  through the HEC Project No.
20-1283.

\vspace{.5in} \small\textbf{Table 1:} Ratios of calculations of
neutrino energy loss rates due to $^{54}Fe$ at different selected
densities and temperatures. $R_{\nu}$ denotes the ratio of the
calculated neutrino energy loss rates to those calculated by large
scale shell model (LSSM) and those calculated by Fuller and
collaborators (FFN).
\begin{center}
\scriptsize\begin{tabular}{ccccccccc}  \hline\hline $T_{9}$ &
$R_{\nu}$(LSSM) & $R_{\nu}$(FFN) &$R_{\nu}$(LSSM) & $R_{\nu}$(FFN)
& $R_{\nu}$(LSSM) & $R_{\nu}$(FFN) & $R_{\nu}$(LSSM) &
$R_{\nu}$(FFN)
\\\hline
& $10gcm^{-3}$ & $10gcm^{-3}$ & $10^{3}gcm^{-3}$ & $10^{3}gcm^{-3}$
&
        $10^{7}gcm^{-3}$ & $10^{7}gcm^{-3}$ & $10^{11}gcm^{-3}$& $10^{11}gcm^{-3}$\\\hline

1  & 4.68E+00&    9.66E-02 &   4.69E+00&   9.66E-02 &   3.96E+00& 1.13E-01&   8.32E-01&   2.96E-01\\
3 &  2.37E+00&    1.48E-01 &   2.37E+00 & 1.48E-01 &   2.47E+00& 1.71E-01 &   8.30E-01&   2.95E-01\\
10&  8.15E-01&    3.90E-01  &  8.17E-01 & 3.91E-01&   8.34E-01 & 3.94E-01&   8.11E-01&   2.90E-01\\
30&  1.05E+00&    3.60E-01&   1.05E+00 & 3.61E-01&    1.05E+00 &3.61E-01   & 1.01E+00&   3.52E-01\\\hline\hline

\end{tabular}
\end{center}

\vspace{.5in} \small\textbf{Table 2:} Same as Table 1, but for
neutrino energy loss rates due to $^{55}Fe$.
\begin{center}
\scriptsize\begin{tabular}{ccccccccc}  \hline\hline $T_{9}$ &
$R_{\nu}$(LSSM) & $R_{\nu}$(FFN) &$R_{\nu}$(LSSM) & $R_{\nu}$(FFN)
& $R_{\nu}$(LSSM) & $R_{\nu}$(FFN) & $R_{\nu}$(LSSM) &
$R_{\nu}$(FFN)
\\\hline
& $10gcm^{-3}$ & $10gcm^{-3}$ & $10^{3}gcm^{-3}$ &
$10^{3}gcm^{-3}$ &
        $10^{7}gcm^{-3}$ & $10^{7}gcm^{-3}$ & $10^{11}gcm^{-3}$& $10^{11}gcm^{-3}$\\\hline

1 &  1.01E+00 &   9.86E-01  &  1.01E+00 &   9.89E-01 &   1.02E+00 &  1.01E+00 & 8.09E-01 &  1.62E-01\\
3 &  1.52E+00 &   3.74E-01  &  1.52E+00&    3.75E-01 &   1.49E+00&   3.73E-01 & 7.96E-01 &  1.62E-01\\
10&  1.00E+00 &   1.39E-01&   1.00E+00&    1.39-E01&     1.02E+00 &    1.39E-01 & 8.41E-01 &  1.78E-01\\
30&  1.67E+00 &   3.31E-01 &  1.68E+00&     3.32E-01&    1.68E+00 &
                                                                       3.33E-01 & 1.42E+00 &  3.25E-01\\\hline\hline
\end{tabular}
\end{center}

\vspace{.5in} \small\textbf{Table 3:} Same as Table 1, but for
neutrino energy loss rates due to $^{56}Fe$.
\begin{center}
\scriptsize\begin{tabular}{ccccccccc}   \hline\hline $T_{9}$ &
$R_{\nu}$(LSSM) & $R_{\nu}$(FFN) &$R_{\nu}$(LSSM) & $R_{\nu}$(FFN)
& $R_{\nu}$(LSSM) & $R_{\nu}$(FFN) & $R_{\nu}$(LSSM) &
$R_{\nu}$(FFN)
\\\hline
& $10gcm^{-3}$ & $10gcm^{-3}$ & $10^{3}gcm^{-3}$ &
$10^{3}gcm^{-3}$ &
        $10^{7}gcm^{-3}$ & $10^{7}gcm^{-3}$ & $10^{11}gcm^{-3}$& $10^{11}gcm^{-3}$\\\hline
1 &  8.49E-01&    2.61E-01&   8.49E-01&    2.61E-01&   9.75E-01&    3.15E-01 &  1.08E+00&    3.57E-01\\
3 &  1.04E+00 &   2.36E-01&  1.04E+00 &   2.37E-01&   1.08E+00&    2.44E-01&   1.05E+00&    3.51E-01\\
10&  8.75E-01 &   4.57E-01&  8.77E-01 &   4.58E-01&   8.83E-01&    4.59E-01&   9.95E-01&    3.30E-01\\
30 & 1.12E+00 &   4.56E-01&  1.13E+00 &   4.57E-01&   1.13E+00&
                                                                   4.58E-01&   1.26E+00&    4.17E-01\\\hline\hline
\end{tabular}
\end{center}

\vspace{.5in} \small\textbf{Table 4:}  Same as Table 1, but for
antineutrino energy loss rates due to $^{54}Fe$.
\begin{center}
\scriptsize\begin{tabular}{ccccccccc}   \hline\hline $T_{9}$ &
$R_{\bar{\nu}}$(LSSM) & $R_{\bar{\nu}}$(FFN)
&$R_{\bar{\nu}}$(LSSM) & $R_{\bar{\nu}}$(FFN) &
$R_{\bar{\nu}}$(LSSM) & $R_{\bar{\nu}}$(FFN) &
$R_{\bar{\nu}}$(LSSM) & $R_{\bar{\nu}}$(FFN)
\\\hline
& $10gcm^{-3}$ & $10gcm^{-3}$ & $10^{3}gcm^{-3}$ &
$10^{3}gcm^{-3}$ &
        $10^{7}gcm^{-3}$ & $10^{7}gcm^{-3}$ & $10^{11}gcm^{-3}$& $10^{11}gcm^{-3}$\\\hline
1 &  1.50E-04&  1.51E-04& 1.49E-04&  1.51E-04& 8.75E-04&  9.29E-04& --& --   \\
3 &  5.07E-02&  5.42E-02& 5.08E-02&  5.43E-02& 4.33E-02&  2.64E-02& 5.92E-03&   5.98E-03\\
10&  6.25E-01&  5.55E-01& 6.27E-01&  5.56E-01& 6.24E-01&  5.51E-01& 3.48E-01&   4.74E-01\\
30 & 6.03E+00 & 7.19E-01& 6.05E+00&  7.23E-01& 6.07E+00& 7.23E-01&
                                                                    5.89E+00&   7.13E-01\\\hline\hline

\end{tabular}
\end{center}

\vspace{.5in} \small\textbf{Table 5:}  Same as Table 1, but for
antineutrino energy loss rates due to $^{55}Fe$.
\begin{center}
\scriptsize\begin{tabular}{ccccccccc}  \hline\hline $T_{9}$ &
$R_{\bar{\nu}}$(LSSM) & $R_{\bar{\nu}}$(FFN)
&$R_{\bar{\nu}}$(LSSM) & $R_{\bar{\nu}}$(FFN) &
$R_{\bar{\nu}}$(LSSM) & $R_{\bar{\nu}}$(FFN) &
$R_{\bar{\nu}}$(LSSM) & $R_{\bar{\nu}}$(FFN)
\\\hline
& $10gcm^{-3}$ & $10gcm^{-3}$ & $10^{3}gcm^{-3}$ &
$10^{3}gcm^{-3}$ &
        $10^{7}gcm^{-3}$ & $10^{7}gcm^{-3}$ & $10^{11}gcm^{-3}$& $10^{11}gcm^{-3}$\\\hline
1 &  6.87E-03&    4.98E-03&  6.92E-03&  4.57E-03&  3.15E-03& 2.09E-04& -- & --  \\
3 &  3.69E-01&    2.98E-02&  3.70E-01&  2.99E-02&  2.89E-02& 2.35E-02& 5.11E-04&  2.28E-02\\
10&  8.59E-01&    9.77E-02&  8.61E-01&  9.79E-02&  8.38E-01& 9.75E-02& 2.13E-01&  9.62E-02\\
30&  9.40E+00&    7.18E-01&  9.44E+00&  7.21E-01&  9.42E+00&
                                                             7.21E-01& 9.04E+00&  7.13E-01\\\hline\hline
\end{tabular}
\end{center}
\vspace{.5in} \small\textbf{Table 6:}  Same as Table 1, but for
antineutrino energy loss rates due to $^{56}Fe$.
\begin{center}
\scriptsize\begin{tabular}{ccccccccc} \hline\hline $T_{9}$ &
$R_{\bar{\nu}}$(LSSM) & $R_{\bar{\nu}}$(FFN)
&$R_{\bar{\nu}}$(LSSM) & $R_{\bar{\nu}}$(FFN) &
$R_{\bar{\nu}}$(LSSM) & $R_{\bar{\nu}}$(FFN) &
$R_{\bar{\nu}}$(LSSM) & $R_{\bar{\nu}}$(FFN)
\\\hline
& $10gcm^{-3}$ & $10gcm^{-3}$ & $10^{3}gcm^{-3}$ &
$10^{3}gcm^{-3}$ &
        $10^{7}gcm^{-3}$ & $10^{7}gcm^{-3}$ & $10^{11}gcm^{-3}$& $10^{11}gcm^{-3}$\\\hline

1 &  1.42E-01&    2.09E-03 &  1.42E-01 &  2.07E-03&  4.48E-02&  1.63E-04& --  & --   \\
3 &  1.89E+00&    6.38E-02 &  1.89E+00 &  6.40E-02&  1.66E-01&  1.87E-02&   8.69E-03&   1.43E-01\\
10&  1.15E+00&    7.94E-02 &  1.15E+00&   7.96E-02&  1.10E+00&  7.96E-02&   2.27E-01&   1.04E-01\\
30&  6.50E+00&    3.93E-01 &  6.53E+00&   3.94E-01&  6.53E+00&
                                                                3.94E-01&   6.22E+00&   3.89E-01\\\hline\hline
\end{tabular}
\end{center}

\newpage


\begin{figure}[htbp]
\caption{Cumulative sum of the B(GT$_{-}$) values for $^{54}$Fe. The
energy scale refers to excitation energies in daughter $^{54}$Co.}
\begin{center}
\begin{tabular}{c}
\includegraphics[width=0.8\textwidth]{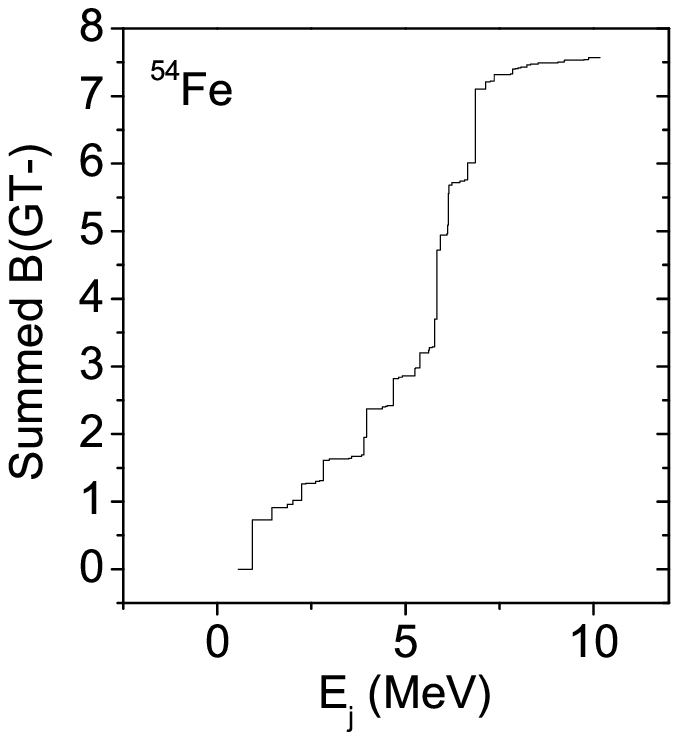}
\end{tabular}
\end{center}
\end{figure}

\begin{figure}[htbp]
\caption{Cumulative sum of the B(GT$_{-}$) values for $^{55}$Fe. The
energy scale refers to excitation energies in daughter $^{55}$Co.}
\begin{center}
\begin{tabular}{c}
\includegraphics[width=0.8\textwidth]{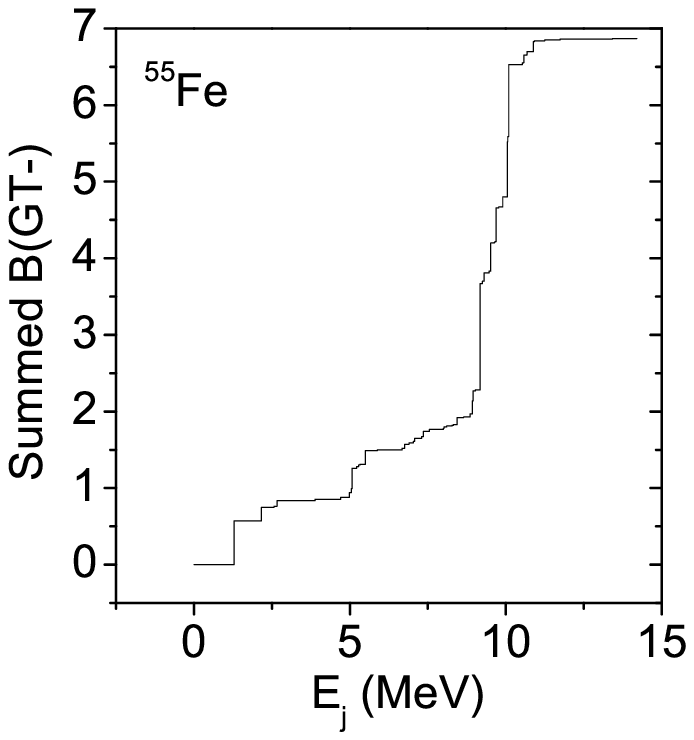}
\end{tabular}
\end{center}
\end{figure}

\begin{figure}[htbp]
\caption{Cumulative sum of the B(GT$_{-}$) values for $^{56}$Fe. The
energy scale refers to excitation energies in daughter $^{56}$Co.}
\begin{center}
\begin{tabular}{c}
\includegraphics[width=0.8\textwidth]{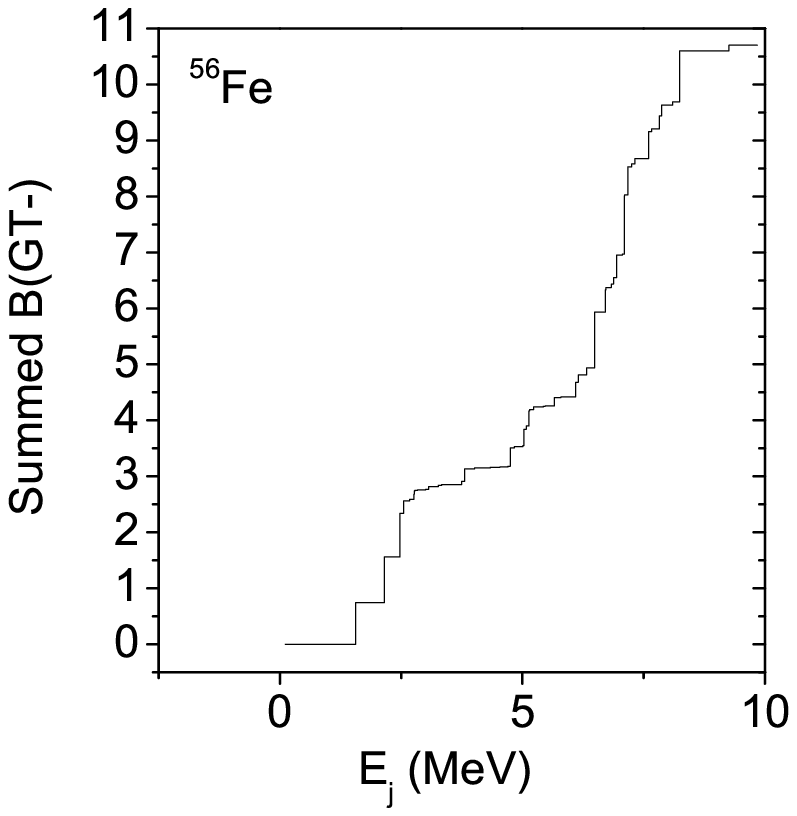}
\end{tabular}
\end{center}
\end{figure}

\begin{figure}[htbp]
\caption{(Color online) Neutrino energy loss rates due to $^{54}$Fe,
as a function of stellar temperatures, for different selected
densities . Densities are in units of $gcm^{-3}$. Temperatures are
given in $10^{9}$ K and log$_{10}\lambda_{\nu}$ represents the log
(to base 10) of neutrino energy loss rates in units of
$MeV.sec^{-1}$.}
\begin{center}
\begin{tabular}{c}
\includegraphics[width=0.8\textwidth]{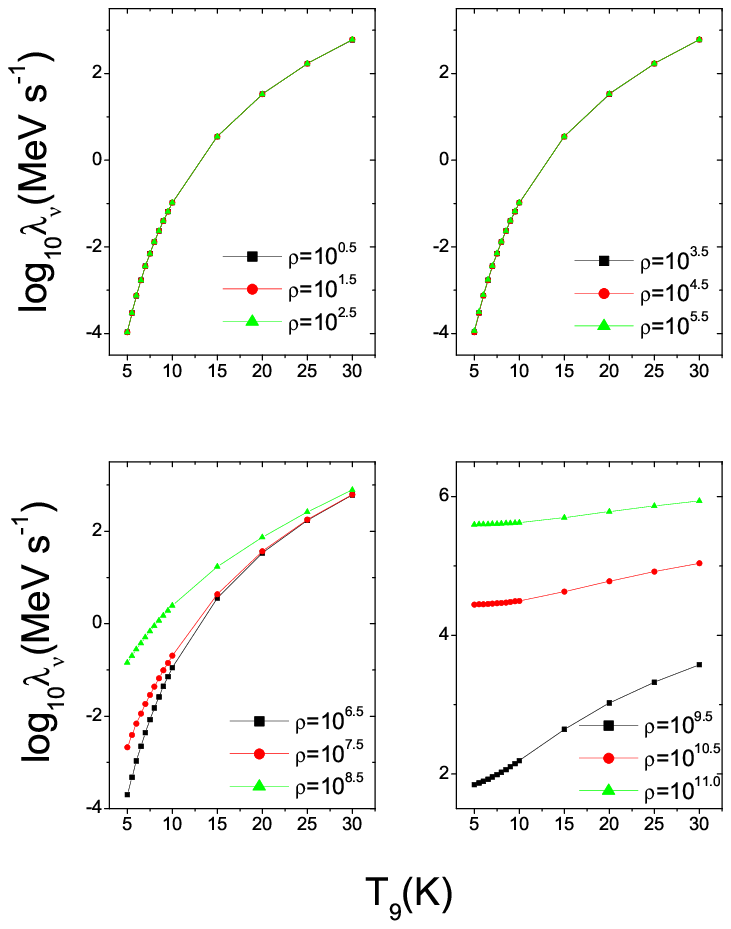}
\end{tabular}
\end{center}
\end{figure}

\begin{figure}[htbp]
\caption{(Color online) Neutrino energy loss rates due to $^{55}$Fe,
as a function of stellar temperatures, for different selected
densities . Densities are in units of $gcm^{-3}$. Temperatures are
given in $10^{9}$ K and log$_{10}\lambda_{\nu}$ represents the log
(to base 10) of neutrino energy loss rates in units of
$MeV.sec^{-1}$.}
\begin{center}
\begin{tabular}{c}
\includegraphics[width=0.8\textwidth]{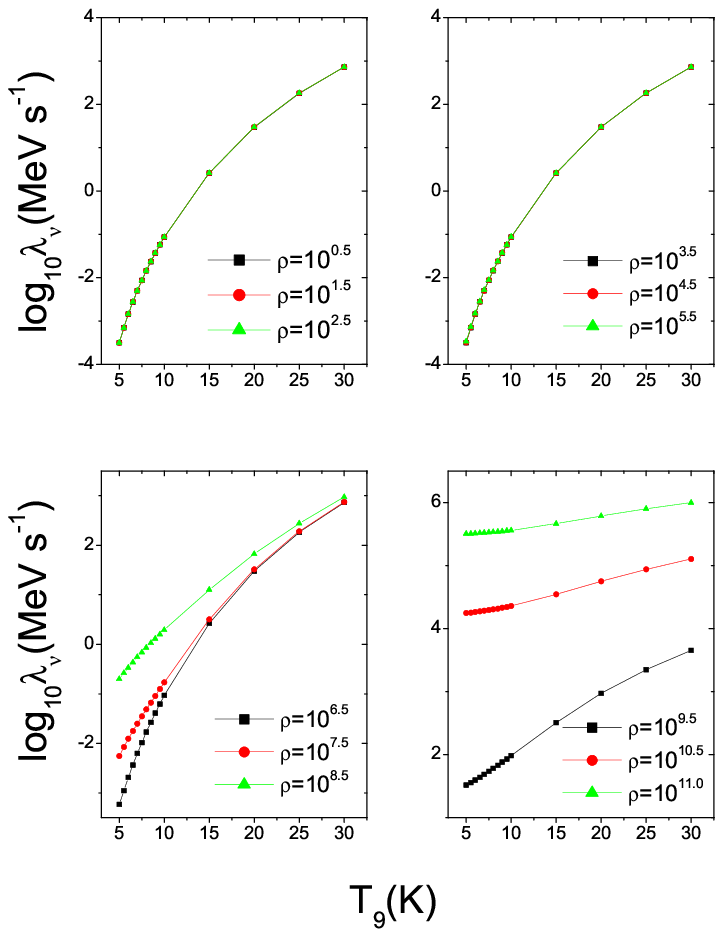}
\end{tabular}
\end{center}
\end{figure}

\begin{figure}[htbp]
\caption{(Color online) Neutrino energy loss rates due to $^{56}$Fe,
as a function of stellar temperatures, for different selected
densities . Densities are in units of $gcm^{-3}$. Temperatures are
given in $10^{9}$ K and log$_{10}\lambda_{\nu}$ represents the log
(to base 10) of neutrino energy loss rates in units of
$MeV.sec^{-1}$.}
\begin{center}
\begin{tabular}{c}
\includegraphics[width=0.8\textwidth]{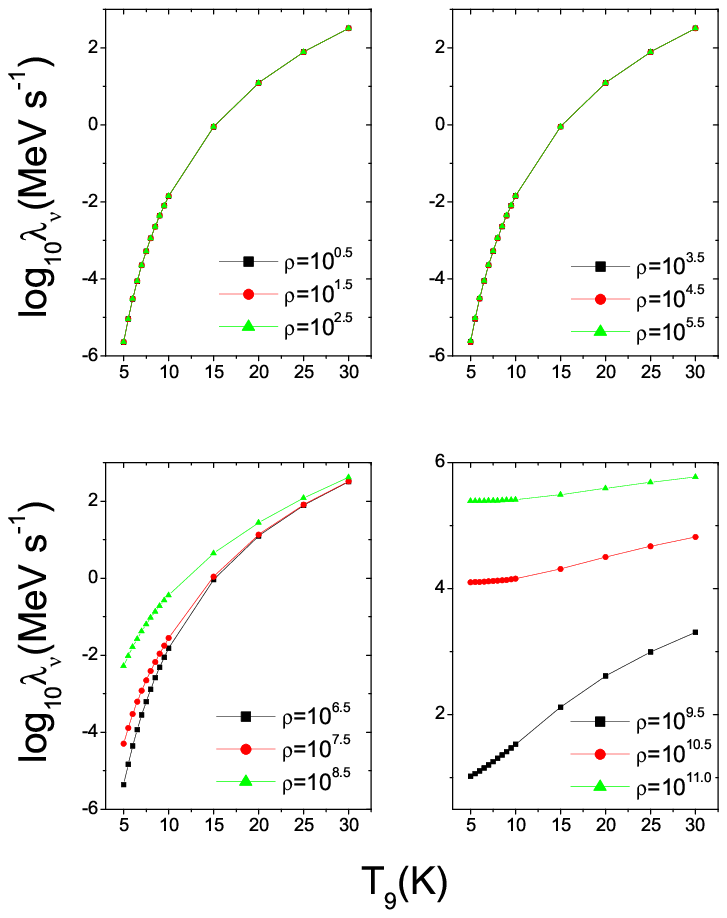}
\end{tabular}
\end{center}
\end{figure}

\begin{figure}[htbp]
\caption{(Color online) Antineutrino energy loss rates due to
$^{54}$Fe, as a function of stellar temperatures, for different
selected densities . Densities are in units of $gcm^{-3}$.
Temperatures are given in $10^{9}$ K and
log$_{10}\lambda_{\bar{\nu}}$ represents the log (to base 10) of
antineutrino energy loss rates in units of $MeV.sec^{-1}$.}
\begin{center}
\begin{tabular}{c}
\includegraphics[width=0.8\textwidth]{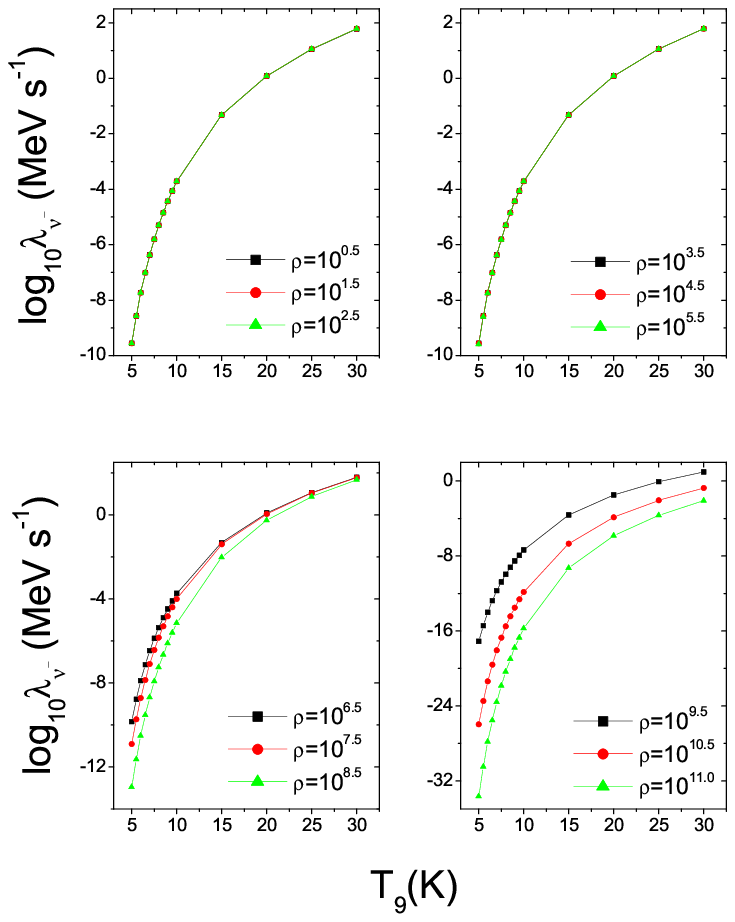}
\end{tabular}
\end{center}
\end{figure}

\begin{figure}[htbp]
\caption{(Color online) Antineutrino energy loss rates due to
$^{55}$Fe, as a function of stellar temperatures, for different
selected densities . Densities are in units of $gcm^{-3}$.
Temperatures are given in $10^{9}$ K and
log$_{10}\lambda_{\bar{\nu}}$ represents the log (to base 10) of
antineutrino energy loss rates in units of $MeV.sec^{-1}$.}
\begin{center}
\begin{tabular}{c}
\includegraphics[width=0.8\textwidth]{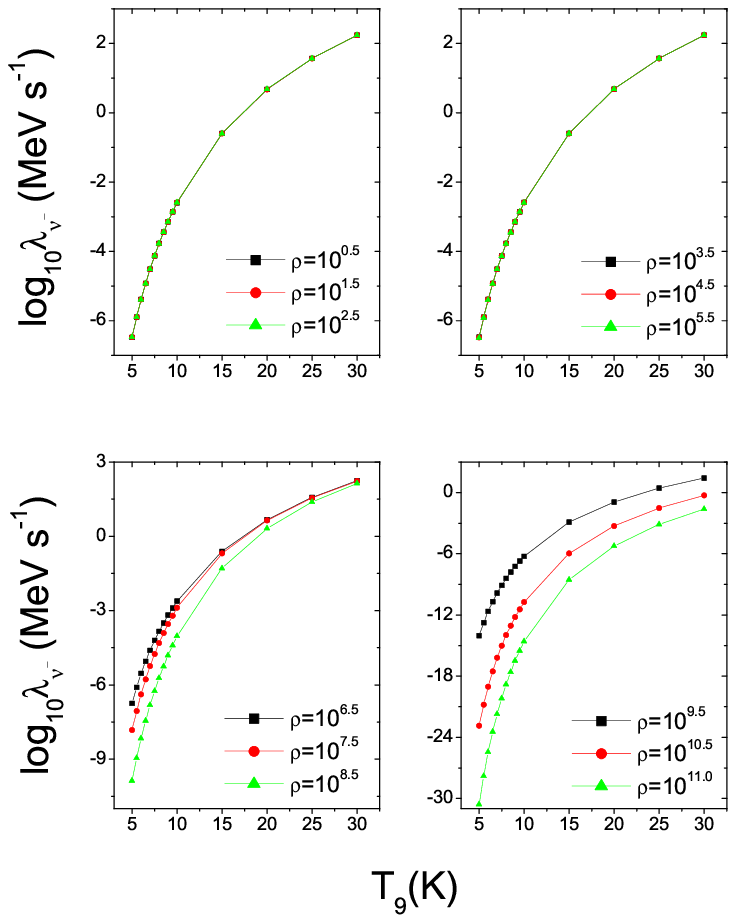}
\end{tabular}
\end{center}
\end{figure}

\begin{figure}[htbp]
\caption{(Color online) Antineutrino energy loss rates due to
$^{56}$Fe, as a function of stellar temperatures, for different
selected densities . Densities are in units of $gcm^{-3}$.
Temperatures are given in $10^{9}$ K and
log$_{10}\lambda_{\bar{\nu}}$ represents the log (to base 10) of
antineutrino energy loss rates in units of $MeV.sec^{-1}$.}
\begin{center}
\begin{tabular}{c}
\includegraphics[width=0.8\textwidth]{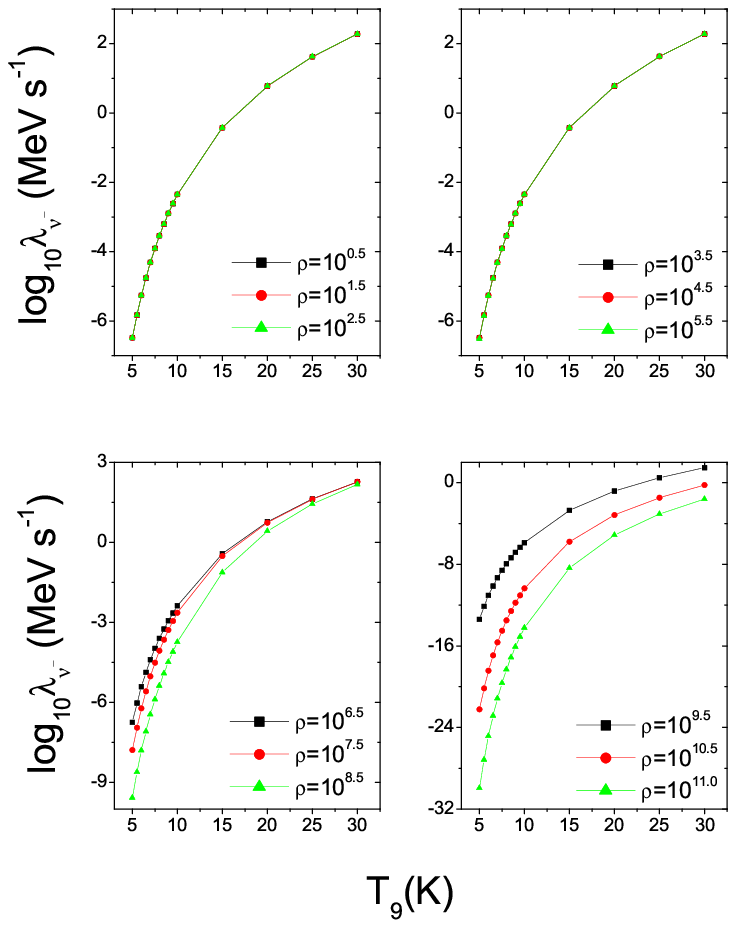}
\end{tabular}
\end{center}
\end{figure}

\begin{thebibliography}{}
\bibitem{Col66}Colgate S. A. \& White R. , The hydrodynamic behavior of supernovae explosions, {\it
Astrophys. J.}, 143, 626-681, 1966.
\bibitem{Arn67}Arnett W. D., Mass dependence in gravitational collapse of stellar cores, {\it Can. J. Phys.}, 45, 1621–1641, 1967.
\bibitem{Bet85}Bethe H. A. \& Wilson J. R. , Revival of a stalled supernova shock by neutrino heating, {\it
Astrophys. J.}, 295, 14-23, 1985.
\bibitem{Bur03}Buras R., Rampp M., Janka H.-T. \& Kifonidis K., Improved models of stellar core collapse and still no explosions: What is missing?,
{\it Phys. Rev. Lett.}, 90, 241101, 2003.
\bibitem{Fry99}Fryer C. L.,  Mass limits for black hole formation, Astrophys. J. {\it
Astrophys. J.}, 522, 413-418, 1999.
\bibitem{Buras06}Buras R., Janka H.-T., Rampp M. \& Kifonidis K., Two-dimensional hydrodynamic core-collapse supernova simulations with spectral neutrino transport - II. Models for different progenitor stars, {\it Astron. Astrophys.}, 457,  281, 2006.
\bibitem{Bur06} Burrows A., Livne E., Dessart L. \& Ott C. D., Multidimensional radiation/hydrodynamic simulations of proto-neutron star convection, {\it Astron. Astrophys.}, 645,  534-550, 2006.
\bibitem{Woo07} Woosley S. E. \& Heger A., Nucleosynthesis and remnants in massive stars of solar metallicity, {\it Phys. Rep.}, 442, 269-283, 2007.
\bibitem{Hax88}Haxton W. C., Neutrino heating in supernovae, {\it Phys. Rev. Lett.}, 60, 1999-2002, 1988.
\bibitem{Kot04}Kotake K., Sawai H., Yamada S. \& Sato K., Magnetorotational effects on anisotropic neutrino emission and convection in core-collapse supernovae, {\it Astrophys. J.}, 608, 391-404, 2004.
\bibitem{Wal05}Walder R., Burrows A., Ott C. D., Livne E., Lichtenstadt I. \& Jarrah M., Anisotropies in the neutrino fluxes and heating profiles in two-dimensional, time-dependent, multigroup radiation hydrodynamics simulations of rotating core-collapse supernovae, {\it Astrophys. J.}, 626, 317-332, 2005.
\bibitem{Esp03}Esposito S., Mangano G., Miele G., Picardi I. \& Pisanti O.,  Neutrino energy loss rate in a stellar plasma, {\it Nuc. Phys. B}, 658, 217-253, 2003.
\bibitem{Ful80} Fuller G. M., Fowler W. A. \& Newman M. J.,
Stellar Weak-Interaction Rates for sd-Shell Nuclei. I. Nuclear
Matrix Element Systematics with Application to $^{26}$Al and
Selected Nuclei of Importance to the Supernova Problem, {\it
Astrophys. J. Suppl.}, 42, 447-473, 1980;  Stellar Weak Interaction
Rates for Intermediate Mass Nuclei. II. A = 21 to A = 60, {\it
Astrophys. J.}, 252, 715-740, 1982; Stellar Weak Interaction Rates
for Intermediate Mass Nuclei. III. Rate Tables for the Free Nucleons
and Nuclei with A = 21 to A = 60, {\it Astrophys. J. Suppl.}, 48,
279-320, 1982; Stellar Weak Interaction Rates for Intermediate Mass
Nuclei. IV. Interpolation Procedures for Rapidly Varying Lepton
Capture Rates Using Effective log (ft)- Values, {\it Astrophys. J.},
293, 1-16, 1985.
\bibitem{Auf94} Aufderheide M. B., Fushiki I., Woosley S. E., Stanford
E. \& Hartmann D. H., Search for Important Weak Interaction Nuclei
in Presupernova Evolution, {\it Astrophys. J. Suppl. Ser}, 91,
389-417, 1994.
\bibitem{Lan00} Langanke K. \& Mart\'{i}nez-Pinedo G., Shell-Model
Calculations of Stellar Weak Interaction Rates: II. Weak Rates for
Nuclei in the Mass Range A = 45-65 in Supernovae Environments, {\it
Nucl. Phys.}, A673, 481-508, 2000.
\bibitem{Nab04} Nabi J.-Un \& Klapdor-Kleingrothaus H. V., Microscopic
Calculations of Stellar Weak Interaction Rates and Energy Losses for
fp- and fpg-Shell Nuclei, {\it At. Data Nucl. Data Tables}, 88,
237-476, 2004.
\bibitem{Joh92}Johnson C. W., Koonin S. E., Lang G. H. \& Ormand W. E., Monte-carlo methods for the nuclear shell-model,  {\it Phys. Rev. Lett.}, 69, 3157-3160, 1992.
\bibitem{Nab09}Nabi J.-Un, Weak-Interaction-Mediated Rates on
Iron Isotopes for Presupernova Evolution of Massive Stars, {\it Eur.
Phys. J. A}, 40, 223-230, 2009.
\bibitem{Nab10}Nabi J.-Un, Stellar $\beta^{\pm}$ decay rates
of iron isotopes and its implications in astrophysics, {\it Adv.
Space Res.}, doi:10.1016/j.asr.2010.05.026, 2010.
\bibitem{Heg01} Heger A., Woosley S. E., Mart\'{i}nez-Pinedo G. \&
Langanke K.,  Presupernova Evolution with Improved Rates for Weak
Interactions, {\it Astrophys. J.}, 560, 307-325, 2001.
\bibitem{Kar94}Kar K., Ray R. \& Sarkar S., Beta-decay rates of fp shell nuclei with
a-greater-than-60 in massive stars at the presupernova stage, {\it
Astrophys. J.}, 434, 662-483, 1994.
\bibitem{Nab99} Nabi J.-Un \& Klapdor-Kleingrothaus H. V., Microscopic
Calculations of Weak Interaction Rates of Nuclei in Stellar
Environment for A = 18 to 100, {\it Eur. Phys. J. A}, 5, 337-339,
1999.

\bibitem{Gov71}Gove N. B. \& Martin M. J., Log-f Tables for Beta
decay, {\it At. Data Nucl. Data Tables}, 10, 205-317, 1971.

\bibitem{Nab99a} Nabi J.-Un \& Klapdor-Kleingrothaus H. V., Weak
Interaction Rates of sd-Shell Nuclei in Stellar Environments
Calculated in the Proton-Neutron Quasiparticle Random-Phase
Approximation, {\it At. Data Nucl. Data Tables}, 71, 149-345, 1999.
\bibitem{Pru03} Pruet J. \& Fuller G. M., Estimates of stellar weak
interaction rates for nuclei in the mass range A = 65--80, {\it
Astrophys. J. Suppl.}, 149, 189-203, 2003.
\end{thebibliography}
\end{document}